\documentclass[superscriptaddress]{revtex4}
\usepackage{graphicx,amssymb,epsfig,psfrag}
\usepackage{amsmath}
\usepackage{amsfonts}
\usepackage{amssymb}
\usepackage{color}
\usepackage{subfigure}

\usepackage[utf8x]{inputenc}
\usepackage{ucs}
\usepackage{mathrsfs}

\newcommand{\tr}[1]{\textrm{#1}}

\newcommand{\zip}[1]{ }
\newcommand*{\nh}{\mathbf{n}}
\newcommand*{\tauh}{\boldsymbol{\tau}}

\newcommand*{\ex}{\mathbf{e}_x}
\newcommand*{\ey}{\mathbf{e}_y}

\newcommand{\pard}[2]{\frac{\partial #1}{\partial #2}}
\newcommand{\pardd}[3]{\frac{\partial #1}{\partial #2 \partial #3}}

\def\@fnsymbol#1{\ifcase#1\or *\or \dagger\or \ddagger\or \mathchar "278\or \mathchar "27B\or \|\or **\or \dagger\dagger \or \ddagger\ddagger \else\@ctrerr\fi\relax}
\long\def\symbolfootnote[#1]#2{\begingroup%
\def\thefootnote{\fnsymbol{footnote}}\footnote[#1]{#2}\endgroup}

\begin{document}
\title{The effect of non-uniform damping on flutter in axial flow and energy harvesting strategies}
\author{Kiran Singh}
\email{kiran.singh@cantab.net}
\altaffiliation[Present address:]{OCCAM, The Mathematical Institute, 24-29 St Giles, OX13LB, Oxford}
\affiliation{Department of Mechanics, LadHyX,\\ {\'E}cole Polytechnique, 91128, Palaiseau, France}
\author{S\'ebastien Michelin}
\email{sebastien.michelin@ladhyx.polytechnique.fr}
\affiliation{Department of Mechanics, LadHyX,\\ {\'E}cole Polytechnique, 91128, Palaiseau, France}
\author{Emmanuel de Langre}
\email{delangre@ladhyx.polytechnique.fr}
\affiliation{Department of Mechanics, LadHyX,\\ {\'E}cole Polytechnique, 91128, Palaiseau, France}


\begin{abstract}
The problem of energy harvesting from flutter instabilities in flexible slender structures in axial flows is considered. In a recent study, we used a reduced order theoretical model of such a system to demonstrate the feasibility for harvesting energy from these structures. Following this preliminary study, we now consider a continuous fluid-structure system. Energy harvesting is modelled as strain-based damping and the slender structure under investigation lies in a moderate fluid loading range, for which {the flexible structure} may be destabilised by damping. The key goal of this work is to {analyse the effect of damping distribution and intensity on the amount of energy harvested by the system}.  The numerical results {indeed} suggest that non-uniform damping distributions may significantly improve the power harvesting capacity of the system. For low damping levels, clustered dampers at the position of peak curvature are shown to be optimal. Conversely for higher damping, harvesters distributed over the whole structure are more effective.
\end{abstract}
\maketitle




\section{Introduction}
\label{sec:intro}
Increasing energy demands motivate the interest in energy harvesting concepts, where the idea is to harness the energy of naturally occurring phenomena. At the scale of kilowatts, concepts include energy harvesting from tidal currents \citep{EH_4.4} and ocean waves \citep{EH_4.6}. At the lower end of the power spectrum, concepts based on photo/thermovoltaics and magneto/piezoelectrics show the scope for powering sensors and mobile electronic devices \citep{MagEH_6p4}; these include energy scavenging from ambient vibrations in structures such as buildings and bridges and oscillatory motion of wheels in automobiles or turbines in engines \citep{EHEC_1p2, EHEC_1p3}. Energy harvesting from fluid-structure interactions (FSI) includes concepts such as vortex induced vibrations (VIV) of bluff bodies in cross-flow \citep{EH_7p9,EH_7p10}, resonant vibrations induced in aerofoils mounted on elastic supports \citep{peng2009,EH_7p6}, flutter of flexible plates \citep{tang2009,EH_5.9, EH_7p1}, and combinations thereof, as for the coupled VIV-flutter energy harvester examined by \cite{EH_7p5}. In this work we focus on harvesting energy from flutter instabilities of slender structures in axial flow.

The classical description of flutter instabilities is self-sustained oscillations that arise due to  the unstable coupling of fluid dynamic pressure and structural bending modes, where for undamped structures the critical speed at flutter onset depends on fluid as well as structural properties \citep{EH_0.2b}. Flutter instabilities are observed in a diversity of configurations, which may be broadly classified as internal or external flows \citep{EH_0.2a,EH_0.2b}. Internal flow instabilities are observed in flexible pipes and channels and are invariably motivated by biological phenomena such as flow induced oscillations in airways and veins \citep{EH_5.11,MagEH_5}. The pipe-conveying fluid is a canonical problem that yields deep insights into FSI; in particular \cite{MagEh_1p8} examined the relationship between local and global instabilities and showed that locally stable configurations may become unstable due to wave reflections at finite boundaries; this destabilising mechanism was recently predicted for compliant channels as well \citep{MagEH_1p4}.  External flow based instabilities include flapping flags \citep{EH_5.5,EH_5.8,eloy2008} and panels \citep{MagEH_6p11} in a steady flow. For all aspect ratios, the plate can become unstable due to fluid loading, defined as the ratio of fluid and structure inertia, and may be represented as a nondimensional length \citep{EH_5.1} or mass ratio \citep{EH_5.6}. \cite{EH_2.1a} and more recently \cite{EH_0.1} examined the occurrence of instabilities in slender structures. They used experimental and theoretical techniques to examine the role of inviscid and viscous drag on static and dynamic instabilities that arise in flexible cylinders in axial flow.  For a recent review on the flutter dynamics of flexible bodies in external flow, the reader is referred to \cite{EH_5.12} and references therein.

In this work, we analyse the scope for harvesting energy in slender elastic cantilevered structures that flutter in a steady flow. From the point of view of the fluid-solid system, energy harvesters are essentially an energy sink, therefore for the theoretical approach adopted here they are modelled as internal damping in the structure. As a first step, we examined the feasibility of this concept using a nonlinear model of a reduced-order system, consisting of a slender cylinder pair connected by discrete springs and dampers \citep{EH_10}. It was shown that the  optimal configuration for this two-degree-of-freedom system is one with  energy harvesters positioned away from the instability source; such a configuration maintains self-sustained flapping in the presence of structural damping.

In this work we generalise this approach for a continuous system, and seek to maximise energy harvesting through carefully-tailored distributions of structural damping. Conventionally, structural instabilities are stabilised by damping \citep{MagEH_5p1,EH_5.11} . However, damping may be destabilising under moderate to heavy fluid loading conditions, as observed in infinite plates \citep{EH_5.2}, flags \citep{EH_5.9} and fluid-conveying pipes \citep{EH_1.3}. \cite{EH_10} show that energy harvesting requires the presence of two traditionally competitive elements -- flutter oscillations and damping. A configuration for which flutter is destabilised by damping is especially interesting from an energy harvesting perspective. This idea has been exploited for  piezoelectric based energy harvesting from flapping flags: \cite{EH_5.9} used linear theory to show the gain in {conversion} efficiency for flags with high fluid loading destabilised by piezoelectric based damping. In this work, we explore this idea using a general model for structural damping (equivalently energy harvesting) and a nonlinear model of the fluid structure interaction of a slender structure in a mean flow.

It is worth noting that existing insights on damping are generally based on the assumption of a constant distribution of damping in the structure {\citep[see for example][]{tang2009}}, whilst little is known about how  nonuniform damping distributions affect the dynamics. In this work we seek clarity on the role of damping distribution on the flutter response of slender structures. The specific motivation is to identify physical mechanisms that maximise this dissipated {(i.e. harvested)} power.

This paper is organised as follows: in Section~\ref{sec:sec2}, the model used for the dynamics of slender structures with non-uniform damping in an axial flow is presented. In Section~\ref{sec:sec3}, the case of uniform damping is considered as a reference configuration and the role of fluid loading on destabilisation by damping is discussed. Computations are then performed for a neutrally buoyant slender cylinder with moderate fluid loading, and the role of damping on the harvested power and  flutter response is examined. Section~\ref{sec:sec4} investigates non-uniform damping distributions  and seeks optimals on two different families of damping functions, { either }distributed over the whole structure or focused on a particular region. In Section~\ref{sec:sec5}, the impact of damping distribution on the flutter dynamics is investigated further to understand the fundamental difference between optimal configurations at low and intermediate damping.

\section{Fluid-solid model}
\label{sec:sec2}
\begin{figure}
\begin{center}
\mbox{
\begin{tabular}{c}
{\includegraphics[width=.5\textwidth,angle=0]{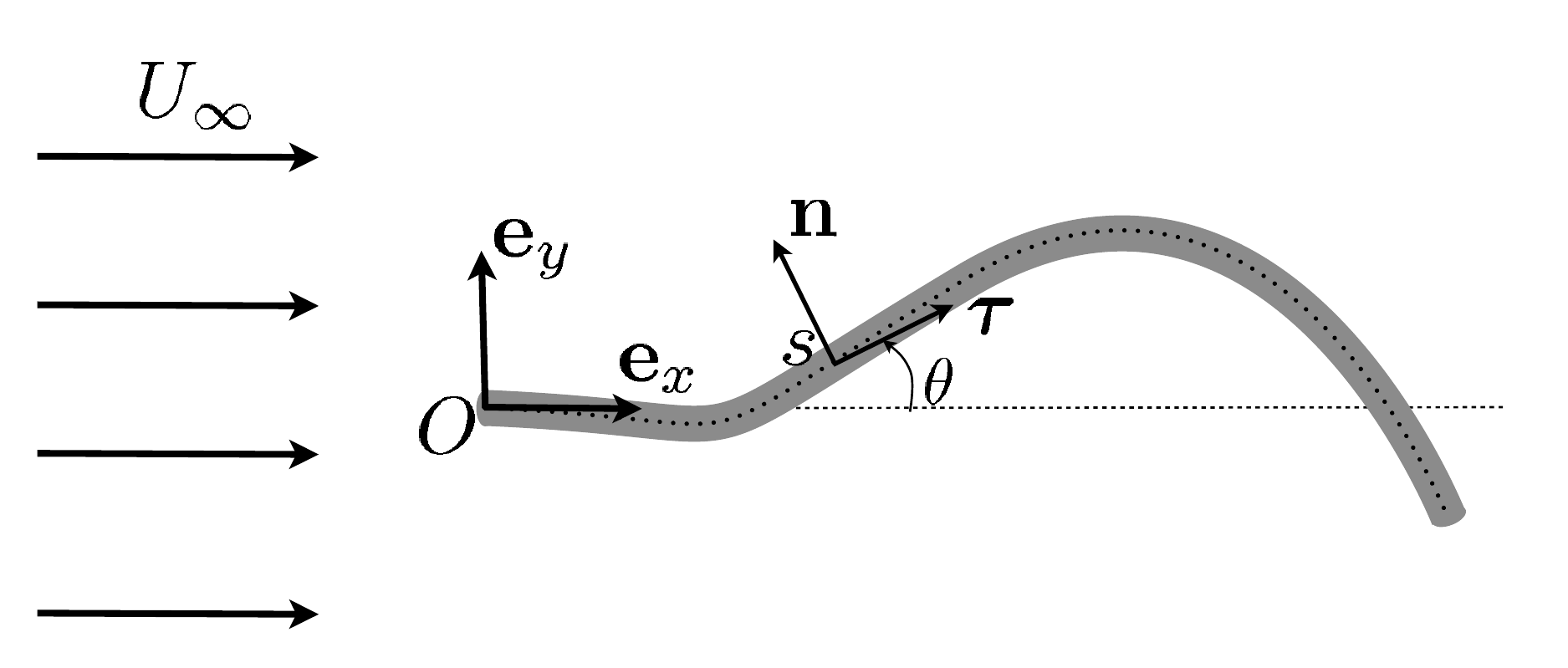}
}
\end{tabular}
}
\end{center}
\caption{Slender cantilevered structure placed in a steady axial flow of velocity $U_\infty \ex$ and fixed in $O$. The instantaneous deformation of the inextensible beam is measured by the tangent angle $\theta(s,t)$ with respect to $\ex$ of the local tangent $\mathbf{\tauh}$.}
\label{fig:sketch}
\end{figure}

We consider a cantilevered (clamped-free, fixed at $O$) slender structure of length $L$ with crosswise dimension $D$, density $\rho_s$, stiffness $B$, and  nonuniform structural damping $B^*(s)$. The slender solid is immersed in a stream of fluid of density $\rho$ moving at mean speed $U_\infty$, and the solid motion is confined to the $(\ex,\ey)$ plane {({Figure~\ref{fig:sketch}})}. The equations of motion are nondimensionalised by the characteristic system scales: $\rho,\ L, U_\infty$. 

\subsection{Nonlinear beam model}
\label{sec:nonlin}

The flexible structure is modelled as an inextensible Euler-Bernoulli beam, where $\mathbf{r}(s)$ is the position vector in the fixed coordinate system $(\ex,\ey)$ and $s$ is the curvilinear coordinate. At each point along the beam, the orientation $\theta(s,t)$ is defined as the angle of the tangent vector $\mathbf{\tauh}(s)$ with the horizontal; $\mathbf{\nh}(s)$ is the local normal. The nonlinear equation of motion for the beam subjected to a fluid force, $\mathbf{f}$, is:
\begin{align}
\frac{1}{M^*}\pard{^2\mathbf{r}}{t^2}=\pard{}{s}\left\{\nu\boldsymbol{\tau}-\frac{1}{M^*{U^*}^2}\left[
\pard{^2\theta}{s^2}+\pard{}{s}\left(\xi(s)\pardd{^2\theta}{s}{t}\right)\right]\mathbf{n}\right\} + \mathbf{f},
\label{eqn:NL_EOM}
\end{align}
where the internal tension, $\nu(s,t)$, is essentially a Lagrange multiplier to satisfy the inextensibility condition $\partial\mathbf{r}/\partial s=\mathbf{\tauh}$ and $\xi(s)=U_\infty B^*(s)/(BL)$ is the non-dimensional damping distribution. {A Kelvin-Voigt damping model is considered here, generalising previous contributions  \citep[e.g.][]{EH_0.2b,tang2009,EH_1.3,EH_5.10} to nonuniform damping distributions.}

The clamped-free boundary conditions must also be satisfied, namely at the fixed end ($s=0$):
\begin{equation}\label{eqn:BC1}
\theta=0,\qquad \mathbf{r}=0,
\end{equation}
and at the free end ($s=1$):
\begin{equation}
\pard{\theta}{s}+\xi\pardd{^2\theta}{s}{t}=0, \quad\pard{^2\theta}{s^2}+\pard{}{s}\left(\xi\pardd{^2\theta}{s}{t}\right)=0,\quad\nu=0.
\label{eqn:BC2}
\end{equation}

Consistently with prior work \citep{ EH_5.6,EH_5.8}, the nondimensional mass ratio $(M^*)$ and flow speed $(U^*)$ are defined as:
\begin{align}
M^*=\frac{\rho D L}{\rho_s A},\qquad U^*=U_\infty L
\bigg(\frac{\rho_s A}{B}\bigg)^{1/2},
\end{align}
with $A$ the cross-sectional area of the structure. 

\subsection{Fluid dynamic model}
\label{sec:fluid}
{In the limit of slender structures ($D\ll L$), and for purely potential flow upstream of the structure's trailing edge, Lighthill's large amplitude elongated-body theory \citep{KSCat1.2} provides a leading order expression for the `reactive' force $\mathbf{f}^i$ applied by the flow on the flapping body, associated with the local transverse motion of each cross-section 
\begin{align}
\mathbf{f}^i= - \frac{m_a}{M^*}\bigg( \frac{\partial (u_n \mathbf{n})}{\partial t}-\frac{\partial (u_n u_\tau \mathbf{n})}{\partial s}+\frac{1}{2}\frac{\partial( u_n^2 \boldsymbol{\tau})}{\partial s} \bigg),
\label{eqn:Lighthill}
\end{align}
where $u_\tau\tauh + u_n \nh = \partial\mathbf{r}/\partial t-\ex$ is the solid's local velocity relative to the incoming flow, and $m_a=M_a/\rho_s A$, where $M_a$ is the dimensional added mass per unit length of the cross-section. \cite{EH_6.4} showed that Lighthill's theory compares well with Reynolds-averaged Navier--Stokes (RANS) simulations to compute the forces on a swimming fish during transient manoeuvres. This reactive force does not include any flow separation associated with the transverse motion of each cross-section, and as emphasised in \cite{EH_6.4}, for freely flapping bodies, an additional contribution for the fluid force must be included to account for such dissipative effects. Here, an empirical `resistive' force model is therefore added following \cite{KSCat2.21} and \cite{EH_5.10}: 
\begin{align}
\mathbf{f}^v=-1/2 C_d u_n |u_n| \mathbf{n},
\label{eqn:Taylor}
\end{align}
where $C_D$ is the empirical drag coefficient, and $C_D=1$ is used in the following for circular cross-sections \citep[see][for a discussion of the impact of this coefficient on the flapping dynamics]{EH_10}. \cite{EH_6.1} tuned the drag coefficients to show good agreement with direct numerical simulations.  }

{
Thus the fluid force, $\mathbf{f}$, in Eq.~\eqref{eqn:NL_EOM} is modelled as the sum of  the reactive ($\mathbf{f}^i$) and resistive ($\mathbf{f}^v$) components. 
In the remainder of the paper we assume a neutrally buoyant circular cylinder, therefore $\rho=\rho_s$ and $A=\pi D^2/4$.  Unless otherwise stated, we assume $D/L=0.1$.}

{Note that the fluid force description is purely local here, and does not explicitly account for wake effects. When the slender body assumption is not verified, and in particular, in the case of two-dimensional plates, an explicit description becomes necessary \citep[see for example][]{EH_5.5,EH_5.8,KSCat10.2}. }

\subsection{Energy harvester model}
As noted earlier, energy harvesting is represented as a strain-based damping $\xi(s)$. {We focus here} on the mean nondimensional harvested power: 
\begin{align}
\mathcal{P}&=\frac{\mathscr{P}}{\rho DLU_\infty^3}=\frac{1}{M^*{U^*}^2}\int_0^1{\xi}(s)\left\langle\dot\kappa^2\right\rangle\ ds, 
\label{eqn:P}
\end{align}
where $\mathscr{P}$ is the mean dimensional harvested power, $\dot\kappa$ is the time derivative of the local curvature $\kappa$ and $\langle\cdot\rangle$ is the time average taken over a period $T$ of the limit-cycle oscillation. $\mathcal{P}$ can also be understood as the efficiency of the system.

As discussed in Section~\ref{sec:intro}, the presence of structural damping can affect the flutter response. {The intensity and  distribution of damping are characterised by}
\begin{align}
\xi_0=\int_0^1\xi(s)\mathrm{d} s, \quad \text{and}\quad \tilde{\xi}(s)=\xi(s)/\xi_0;
\label{eqn:xi_contrnt}
\end{align}
Equation \eqref{eqn:xi_contrnt} allows us to independently evaluate the impact on the system response of (i) the amount of damping $\xi_0$ and (ii) {its spatial distribution}. 

\subsection{Numerical solution}
\label{sec:Numsol}
\begin{figure}
\begin{center}
\mbox{
\begin{tabular}{cc}
\subfigure[]
{\includegraphics[width=.4\textwidth,angle=0]{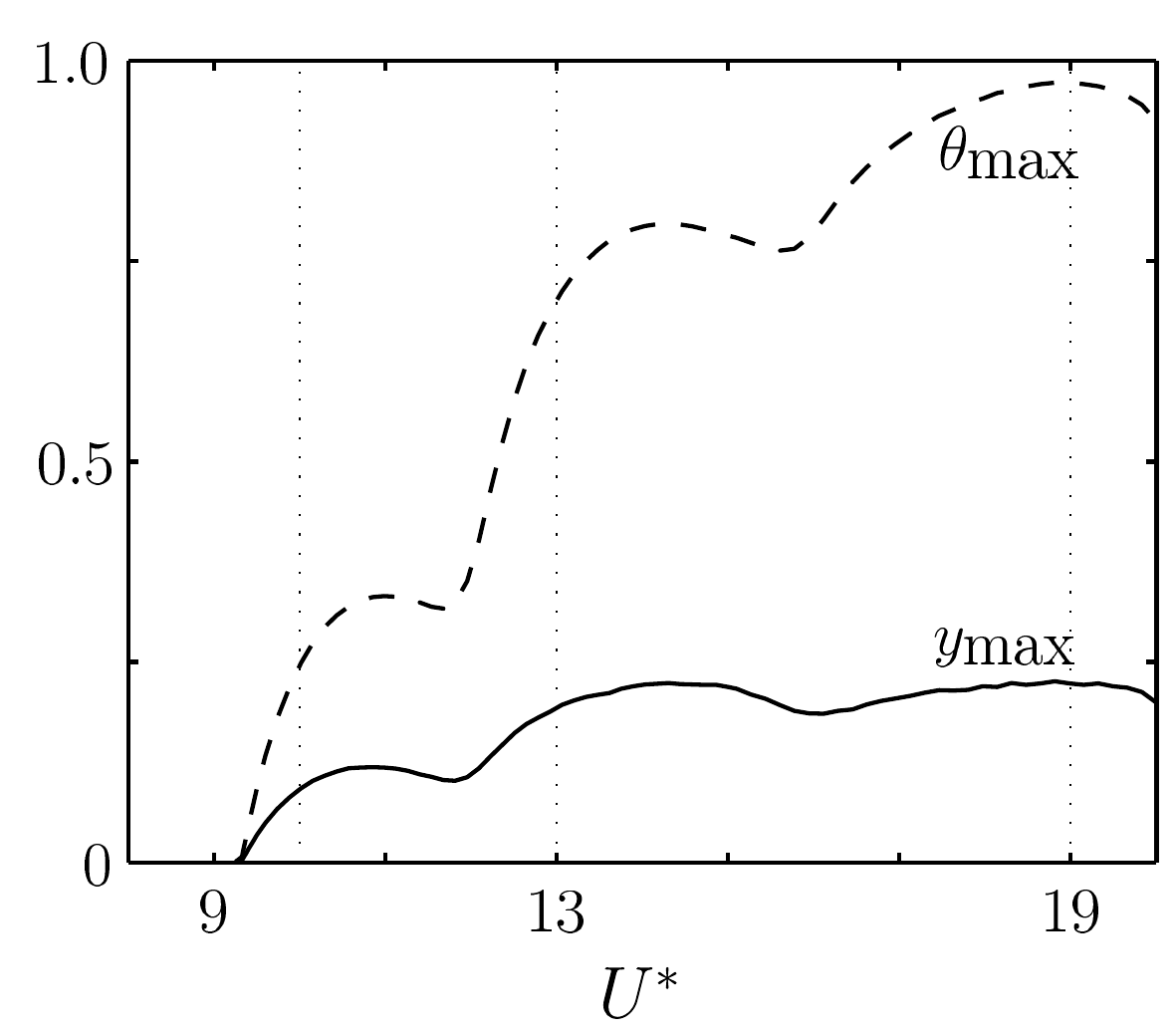}
\label{fig:bifcurve}}
&
\subfigure[]
{\includegraphics[width=.3\textwidth,angle=0]{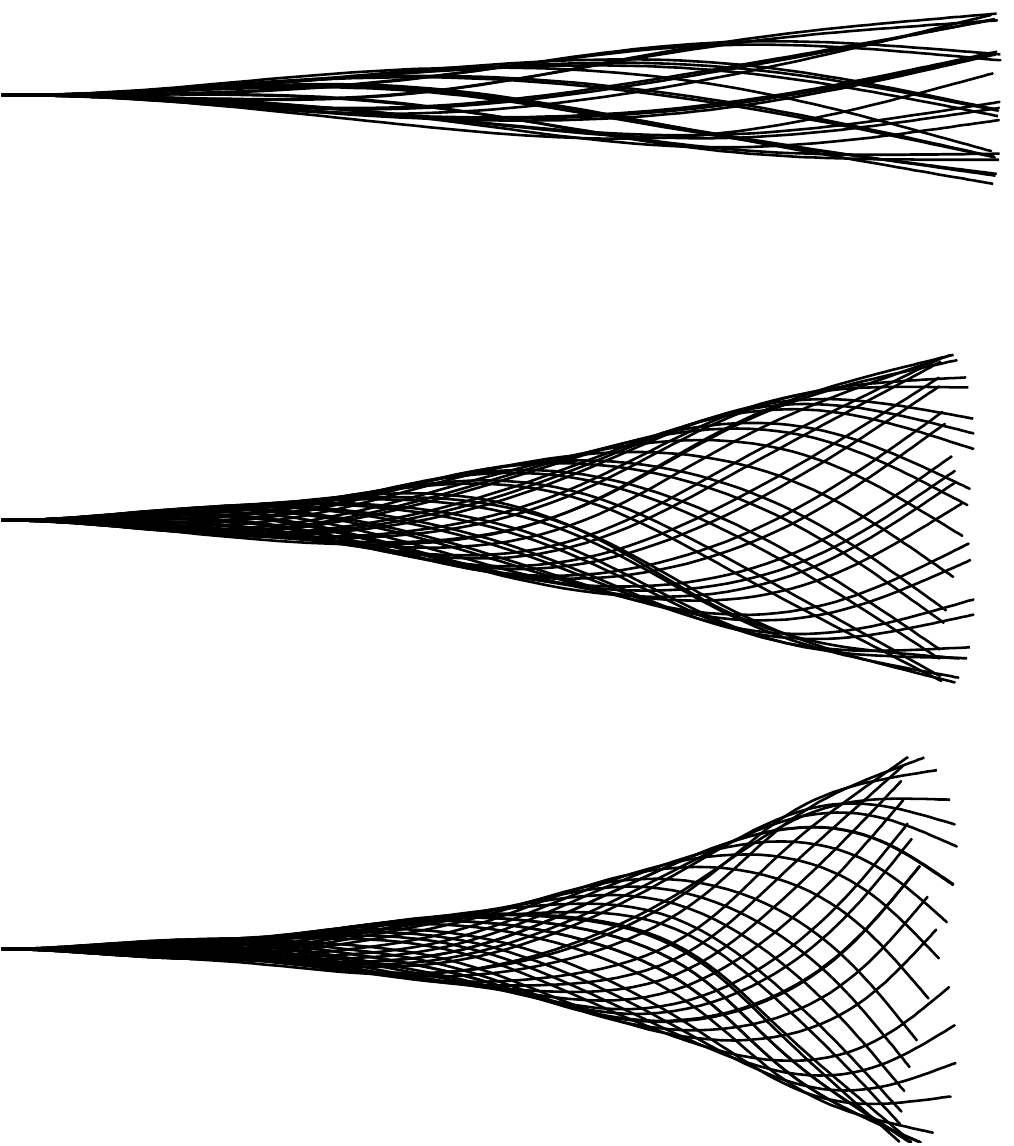}
\label{fig:flutter_ueq9p5}}
\end{tabular}
}
\end{center}
\caption{(a) Maximum deflection $y_\textrm{max}$ (solid) and orientation $\theta_\textrm{max}$ (dashed) at the free end as a function of the non-dimensional flow velocity $U^*$ for $M^*=12.7$. (b) Snapshots of the beam response for $U^*=10$, $13$ and $19$ (from top to bottom).}
\label{fig:sysresp}
\end{figure}

{Equation~\eqref{eqn:NL_EOM} is solved numerically together with boundary conditions \eqref{eqn:BC1}--\eqref{eqn:BC2} using an iterative second-order implicit time-stepping scheme \citep{EH_5.5a}, and spatial derivatives are computed using Chebyshev collocation \citep{Boyd2001}.} 
Conservation of energy is ensured by verifying that $\dot{E}=W_f-Q,$ {where}
\begin{align}
E&=\frac{1}{2}\int_0^1 \left(
\frac{{{|\dot{\mathbf{r}}|}^2}}{M^*}+\frac{\kappa^2}{M^*{U^*}^2}\right) \ ds, \\
Q&=\frac{1}{M^*{U^*}^2}\int_0^1 \xi(s)\dot\kappa^2\ ds, \quad  
W_f=\int_0^1 \mathbf{f}\cdot \dot{\mathbf{r}}\ ds,
\end{align}
{are respectively the mechanical energy of the system, the dissipated power and the rate of work of the fluid forces, and is classically obtained by projecting the equation of motion \eqref{eqn:NL_EOM} on the solid velocity $\dot{\mathbf{r}}$ and by integrating over the entire beam.}
Also note from \eqref{eqn:P}, $\mathcal{P}=\langle Q \rangle$. In the numerical implementation the beam and the flow are initially at rest; the flow speed is ramped up to its steady state value and a small perturbation is applied to the vertical flow. \\
\ \\
\subsection{Nonlinear response of an undamped beam}
Prior to analysing the energy harvesting properties of the system, we first examine the undamped flutter response of the structure. In Figure~\ref{fig:sysresp}, we plot the system response for increasing nondimensional flow speed, for a circular cylinder $(M^* \approx 12.7)$. The critical flow speed at which flutter ensues is verified from linear stability analysis and is confirmed with \cite{EH_2.1a}. Consistent with flutter in plates \citep{EH_5.10}, we note that this instability is a supercritical Hopf bifurcation with flow speed. The jumps in the bifurcation curve correspond to the mode switching reported in \cite{EH_2.1c}, this may also be discerned from the snapshots of the beam at three different flow speeds. For the energy harvesting computations performed in the rest of the paper we set the flow speed at $U^*=13$, corresponding to well-developed oscillations and moderate deflections for the undamped configuration.

\section{Uniform damping}
\label{sec:sec3}
Here, a uniform damping distribution $\xi(s)=\xi_0$ is considered. The dependence of critical flutter speed on the damping {intensity $\xi_0$ is first analysed for varying values of fluid loading $M^*$}. Based on these results, a moderate fluid loading is selected to study the power harvesting capacity of the configuration.

\subsection{Destabilisation by damping: impact on critical flutter speed}
\label{sec:LSA}
\begin{figure}
\begin{center}
\mbox{
\begin{tabular}{c}
{\includegraphics[width=.45\textwidth,angle=0]{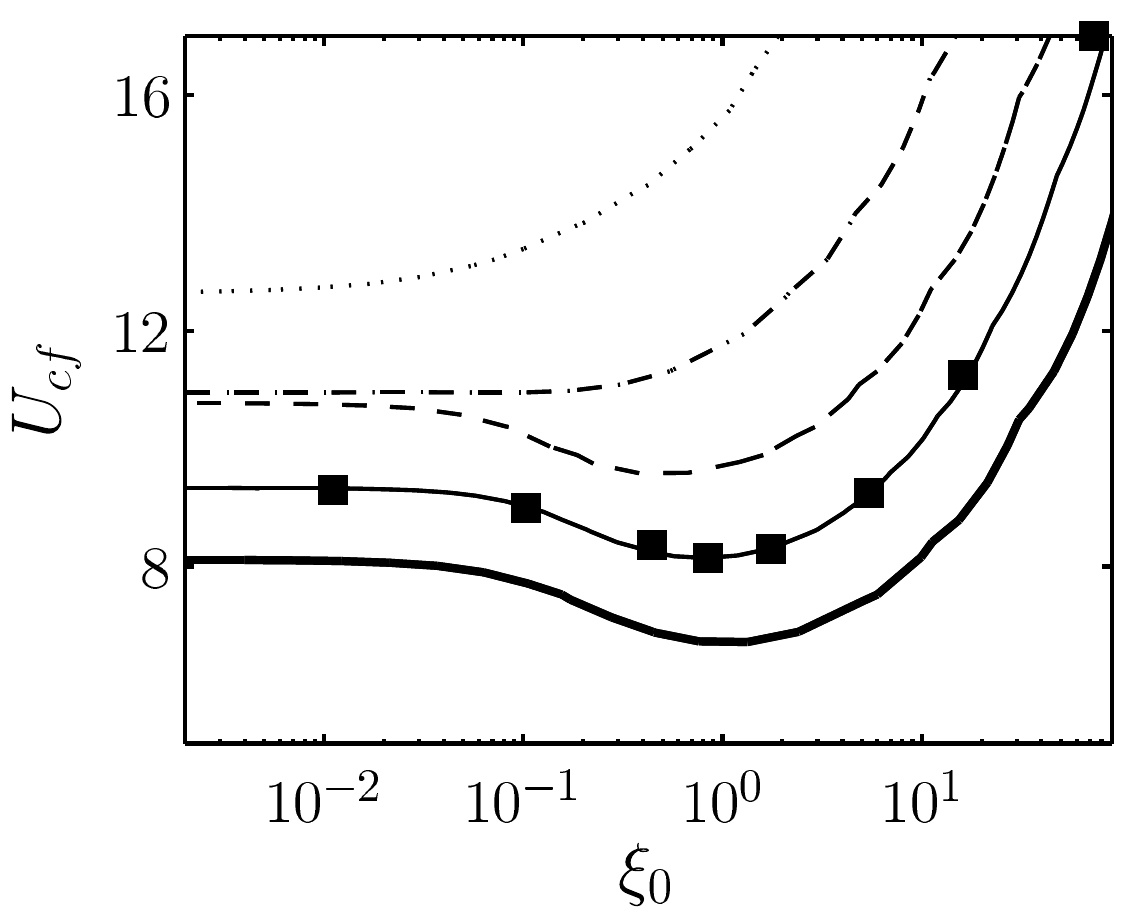}}
\end{tabular}
}
\end{center}
\caption{Variation of the critical flutter speed, $U_\textrm{cf}$ with damping $\xi_0$ for $M^*=2.5$ (dotted), $5.1$ (dash-dotted), $8.5$ (dashed), $12.7$ (solid) and  $20$ (thick solid) obtained from linear stability analysis for uniform damping. Results of nonlinear computations for $M^*=12.7$ are also presented (squares).}
\label{fig:LSA}
\end{figure}

As noted in Section~\ref{sec:intro}, damping can destabilise flexible structures at sufficiently high fluid loading. \cite{EH_1.3} shows that destabilisation of long pipes with a high mass ratio is associated with a drop in the critical flutter speed. Based on this work, we analyse the linear stability of the system and in Figure~\ref{fig:LSA} compare the variation of the critical flutter speed, $U_{cf}$, with damping, $\xi_0$, at different values of fluid loading, $M^*$. For lightly loaded structures ($M^*<5.5$) $U_{cf}$ monotonically increases with damping; at higher values  $(M^*\ge 8.5)$, $U_{cf}$ decreases with damping until a specific value ($\xi_{m}$) above which $U_{cf}$ increases rapidly; note that $\xi_m$ increases
with $M^*$. 

For the remainder of the energy harvesting analysis, we settle on a value of $M^*=12.7$ (corresponding to $D/L=0.1$): this choice allows us to analyse the scope for harvesting energy from a system at moderate fluid loading with destabilisation by damping.

\subsection{Harvesting power with constant damping}
\label{sec:P_const}
\begin{figure}
\begin{center}
\mbox{
\begin{tabular}{c}
{\includegraphics[width=.85\textwidth,angle=0]{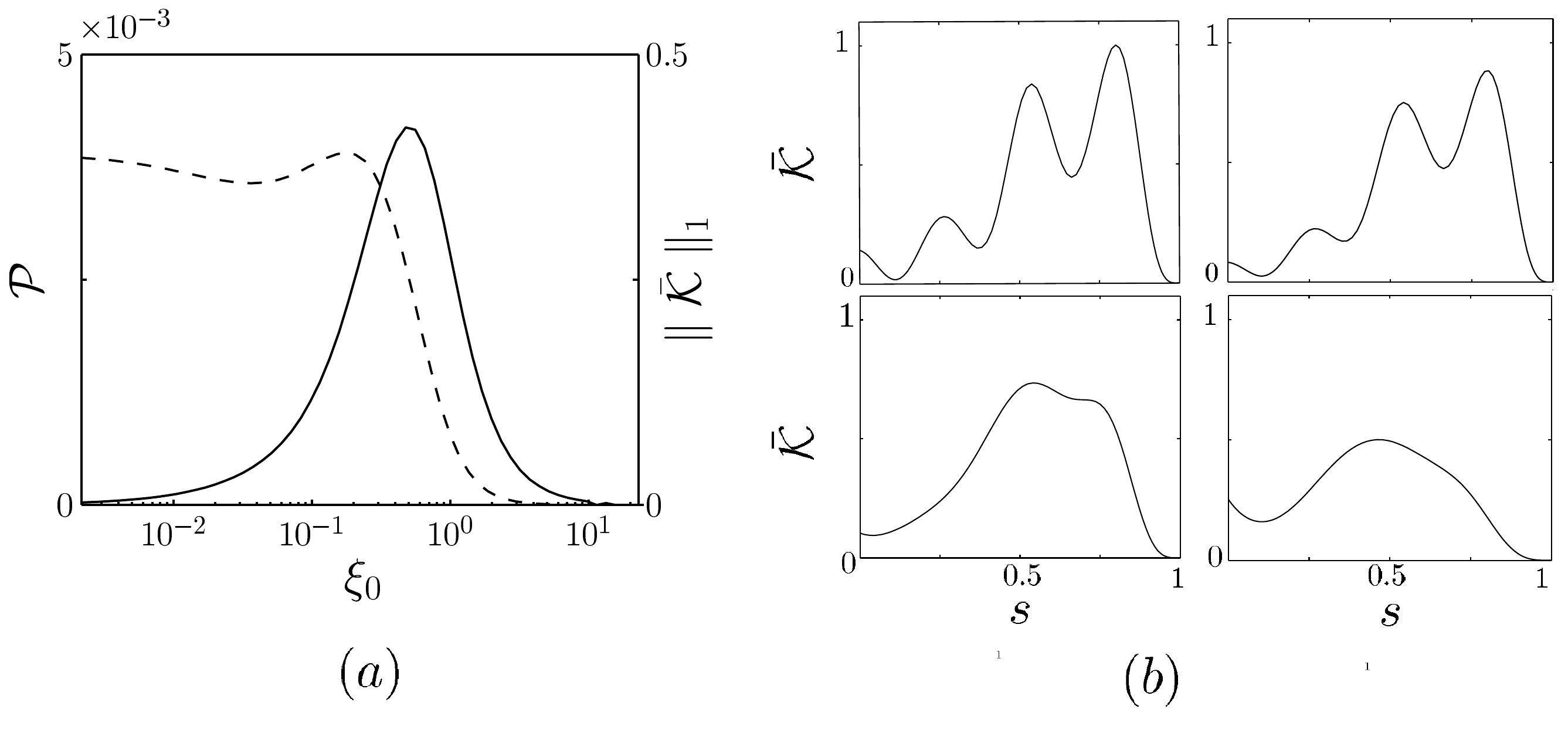}}
\end{tabular}
}
\end{center}
\caption{(a) Evolution of the harvested power $\mathcal{P}$ (solid) and curvature norm $\parallel \bar{\mathcal{K}}\parallel_1$ (dashed)
with the uniform damping intensity $\xi_0$. 
(b) Distribution of curvature-change $\bar{\mathcal{K}}(s)$ along the beam for $\xi_0=0$, $0.022$, $0.22$ and $0.47$ (from left to right and top to bottom, respectively). {($M^*=12.7,\ U^*=13$)}.}
\label{fig:PK_G0}
\end{figure}

Figure~\ref{fig:PK_G0}(a) presents the evolution of harvested power, $\mathcal{P}$,  with damping for $10^{-3}<\xi_0<10$. Equation \eqref{eqn:P} becomes
\begin{align}
\mathcal{P}&=\frac{1}{M^*{U^*}^2}\xi_0\parallel \mathcal{K}\parallel_1, \label{eqn:P_compact} 
\end{align}
where $ \mathcal{K}(s)=\langle \dot\kappa^2 \rangle$ and $\parallel\mathcal{K}\parallel_1=\int_0^1\mathcal{K}\mathrm{d} s$ is the $L_1$-norm of $\mathcal{K}(s)$.
The evolution of the flutter response with damping is examined by plotting the rescaled curvature term, $\bar{\mathcal{K}}=\mathcal{K}/\parallel \mathcal{K}_0\parallel_\infty$ (where $\mathcal{K}_0(s)$ is the value of $\mathcal{K}$ for $\xi_0=0$) for increasing $\xi_0$  (Figure~\ref{fig:PK_G0}b). At small $\xi_0$, the response is virtually no different  from the undamped case, but for $\xi_0 > 0.1$ a perceptible change is observed. First, the curvature is redistributed along the entire beam, and as the damping is increased further, the response of the beam is damped out globally. This can also be seen from the variation of $\parallel\mathcal{K}\parallel_1$ with $\xi_0$ in Figure~\ref{fig:PK_G0}(a). Notably from \eqref{eqn:P_compact} it is $\parallel\mathcal{K}\parallel_1$ that directly impacts the harvested power. 

As a result, for small damping, the change in system response is virtually imperceptible from the undamped case, and power simply scales linearly with $\xi_0$. However, damping has a strong effect on the flutter response at larger $\xi_0$: it reduces the amplitude of curvature-change significantly and causes a sharp reduction in $\mathcal{P}$. The strategy to optimally harvest power is to find the upper bound on $\xi_0$ below which $\parallel \mathcal{K}\parallel_1$ can be maintained close to (or ideally enhanced above) its undamped value.

\section{Nonuniform damping}
\label{sec:sec4}

\begin{figure}[t]
\begin{center}
\mbox{
\begin{tabular}{c}
{\includegraphics[width=.85\textwidth,angle=0]{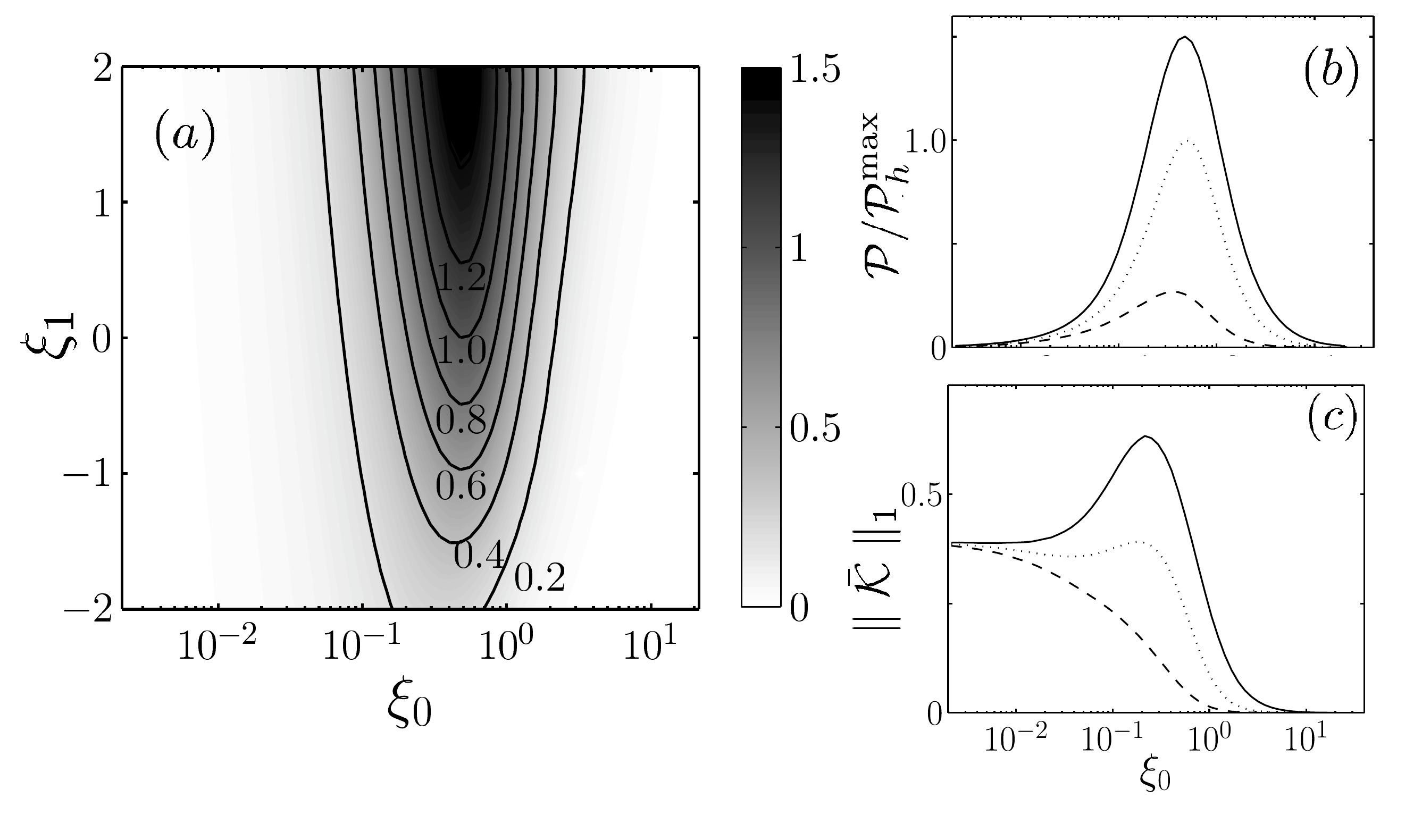}}
\end{tabular}
}
\end{center}
\caption{Linear distribution \eqref{eqn:xi_lin}: (a) rescaled power contours $\mathcal{P}/\mathcal{P}_h^{\tr{max}}$ for varying
  $\xi_0$ and $\xi_1$. (b) Rescaled harvested power $\mathcal{P}/\mathcal{P}_h^{\tr{max}}$ and (c) $\parallel \bar{\mathcal{K}}\parallel_1$ as a function of the damping intensity $\xi_0$ for linearly decreasing ($\xi_1=-2$, dashed), constant ($\xi_1=0$, dotted) and linearly increasing ($\xi_1=2$, solid) damping distributions. {($M^*=12.7,\ U^*=13$)}.  }
\label{fig:LinCon}
\end{figure}

The results for constant damping suggest that as long as $\parallel\mathcal{K}\parallel_1$ is maintained at undamped levels, the harvested power increases linearly with $\xi_0$. Figure \ref{fig:PK_G0}(b) clearly shows that $\mathcal{K}(s)$ varies significantly along the length of the beam. Concentrating harvesters around the zone of peak curvature could therefore enhance the harvested power. This idea is tested in this section on a reduced functional space for the damping distribution $\xi(s)$. {Two families of damping functions are considered, namely (a) dispersed and (b) focused damping distributions. Optimisation of the harvested power is performed within each family, in anticipation of insights into the global optimal}. 

\subsection{Dispersed harvester distribution}
\label{sec:lin}

In this section, we are interested in simple non-homogeneous distributions of damping of the form:
\begin{align}
{\xi}(s)=\xi_0\big (1+\xi_1(s-{1}/{2})\big)
\label{eqn:xi_lin}
\end{align}
characterised by the total damping, $10^{-3}<\xi_0<10$, and slope, $-2\le\xi_1\le2$. This function family corresponds to a dispersed distribution, where the damping is significant over the entire length of the structure. Figure~\ref{fig:LinCon}(a) shows the variation of rescaled power, $\bar{\mathcal{P}}={\mathcal{P}}/\mathcal{P}_h^\textrm{max}$ {with $(\xi_0,\xi_1)$}, where $\mathcal{P}_h^\textrm{max}$ is the maximum harvested power for constant damping (Section~\ref{sec:sec3}). Figure~\ref{fig:LinCon} shows that for all $\xi_0$, the optimal distribution corresponds to $\xi_1=2$, {when damping is distributed increasingly from fixed to free end, and that this linear distribution of damping leads to an increase of the maximum harvested power by $50\%$ compared to the uniform distribution ($\xi_1=0$).} Through non-uniform but simple distributions of damping on the structure, it is indeed possible to enhance the flutter response above that of the undamped configuration (Figure \ref{fig:LinCon}c). 

\subsection{Focused harvester distribution}
\label{sec:gauss}

The results from the reduced order analysis \citep{EH_10} suggest that the optimal distribution ought to be localised at specific points on the beam. The chief drawback of the two parameter functions examined in the previous section is the inability to consider localised distributions of damping in specific regions.  To consider such peaked distributions, we now turn to the following three-parameter family of Gaussian damping distribution: 
\begin{align}
{\xi}(s)=\xi_0
\frac{\xi_g}{\parallel \xi_g \parallel_1},\
\xi_g=\mathrm{e}^{-\alpha(s-s_o)^2},
\label{eqn:Gauss}
\end{align}
where $\xi_0$ is the total damping in the system, $s_o$ is the centre of the distribution and $1/\alpha$ is a measure of its spread in $s$. For increasing $\alpha$ energy harvesters are increasingly concentrated {around $s_o$}.

\begin{figure}[t]
\begin{center}
\mbox{
\begin{tabular}{cc}
\subfigure[]
{\includegraphics[width=.4250\textwidth,angle=0]{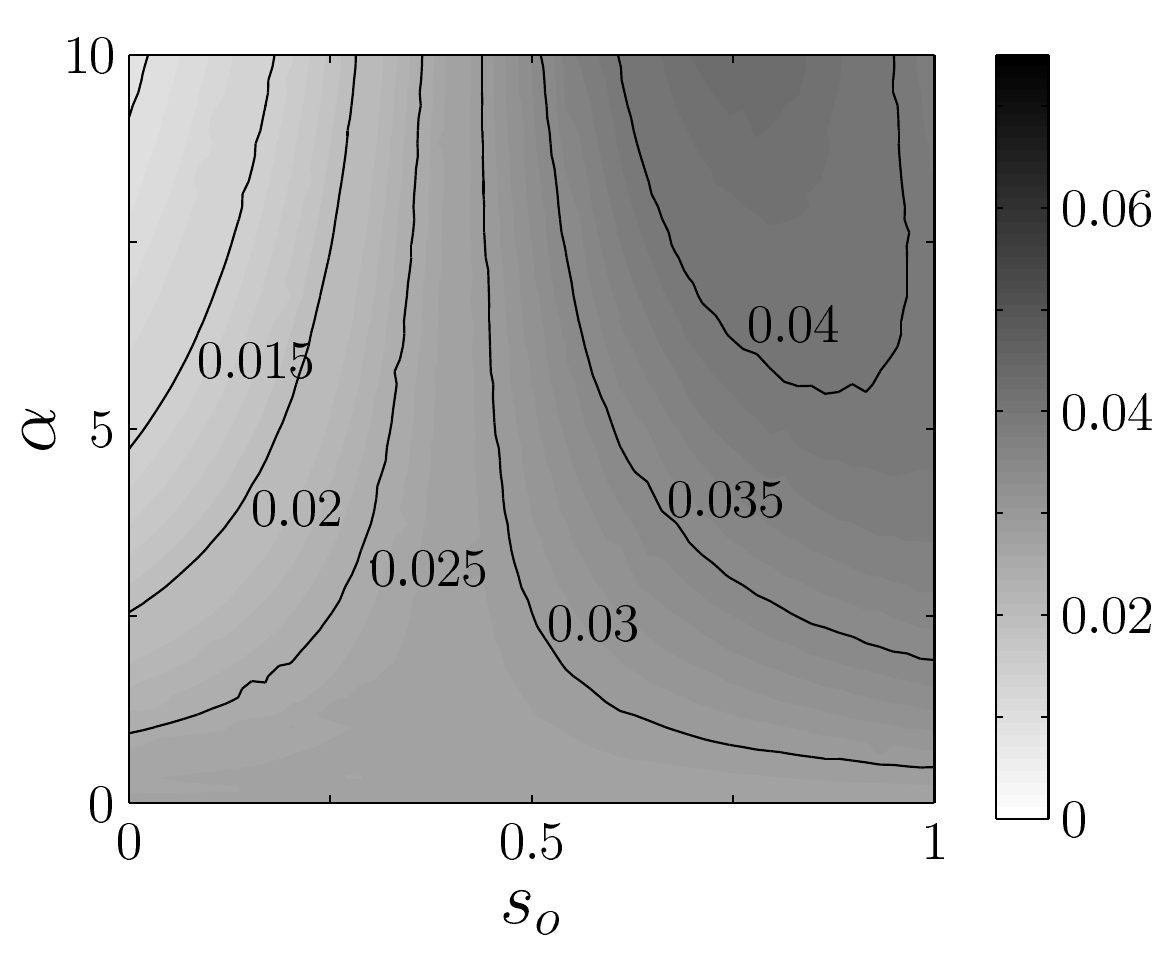}
\label{fig:6a}}
&
\subfigure[]
{\includegraphics[width=.4170\textwidth,angle=0]{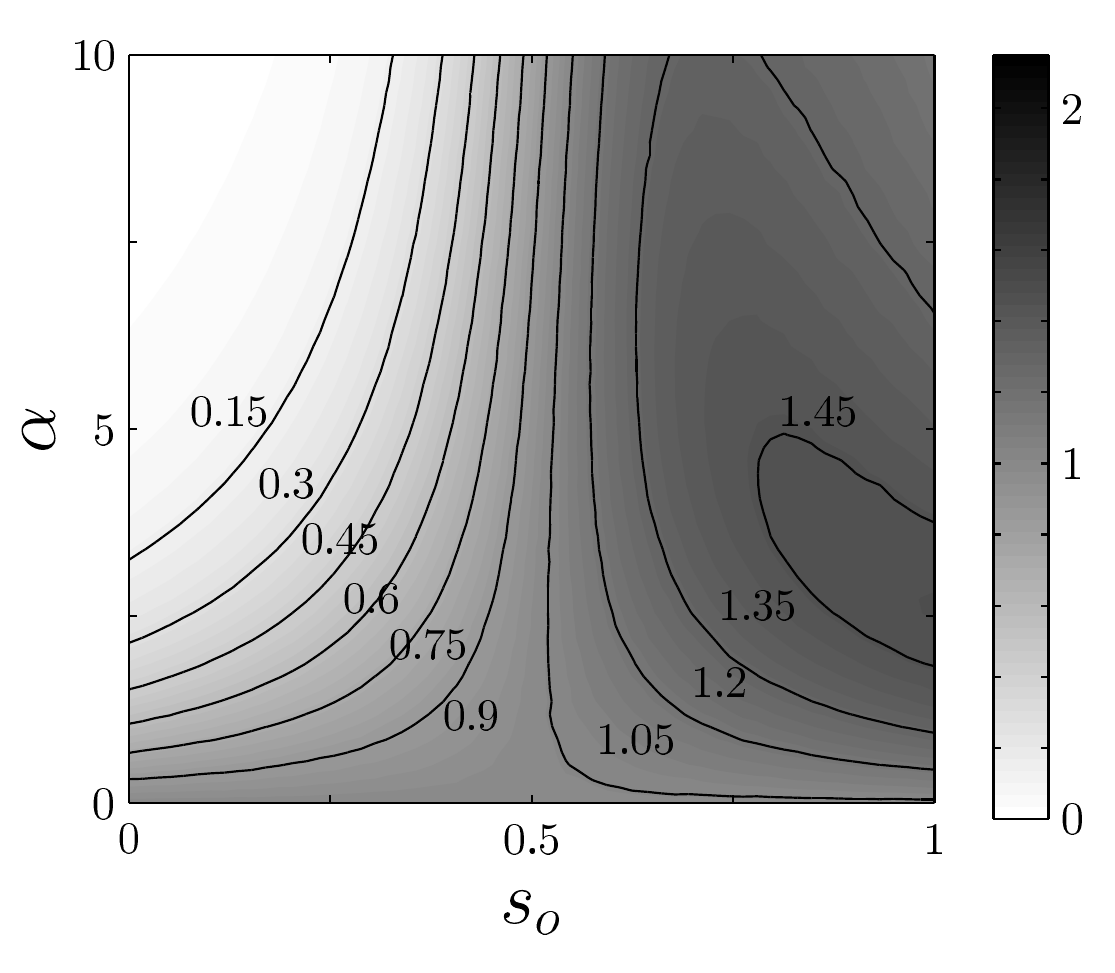}
\label{fig:6b}}
\end{tabular}
}
\end{center}
\caption{Gaussian distribution \eqref{eqn:Gauss}: Maps of rescaled power $\mathcal{P}/\mathcal{P}_h^{\tr{max}}$ for varying $\alpha$ and $s_o$ at (a) $\xi_0=0.01$ and (b) $\xi_0=0.47$. {($M^*=12.7,\ U^*=13$)}.  }
\label{fig:GaussCon}
\end{figure}

\begin{figure}
\begin{center}
\mbox{
\begin{tabular}{cc}
\subfigure[]
{\includegraphics[width=.425\textwidth,angle=0]{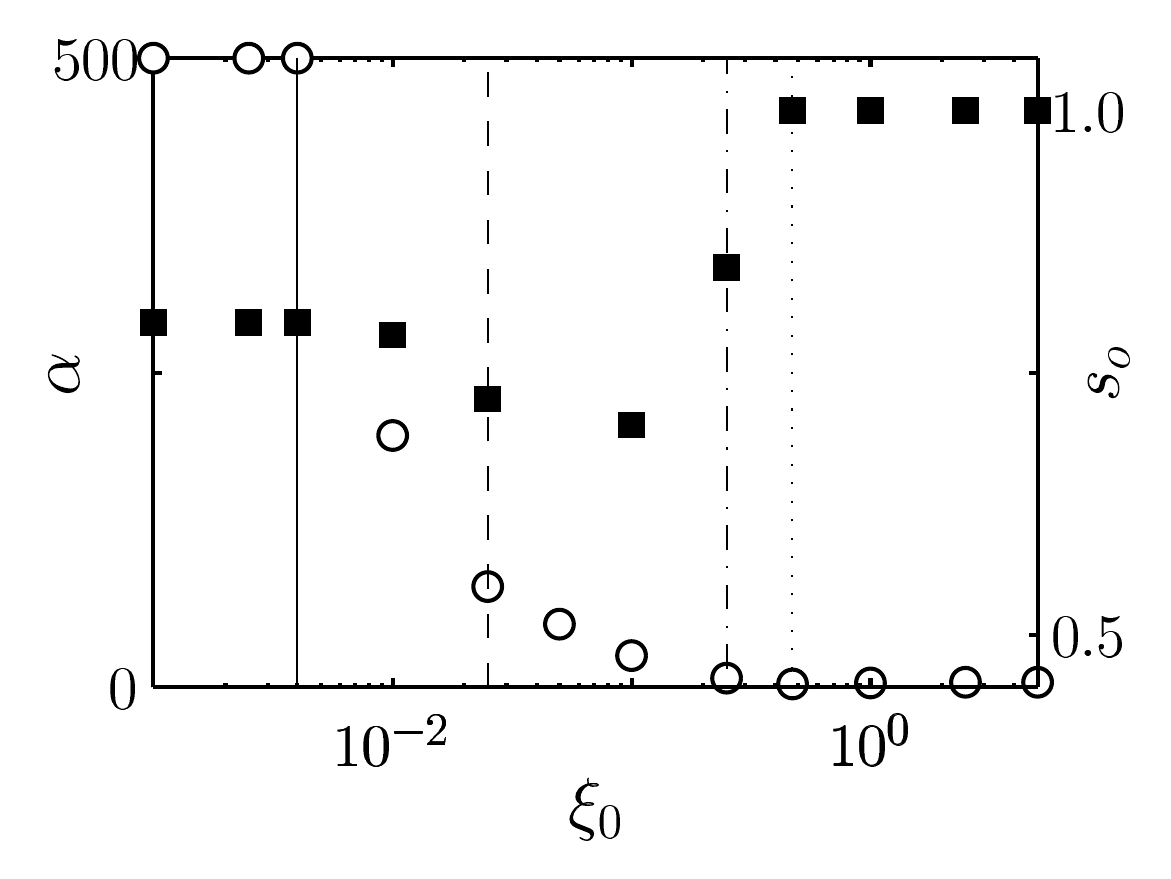}
\label{fig:7a}}
&
\subfigure[]
{\includegraphics[width=.405\textwidth,angle=0]{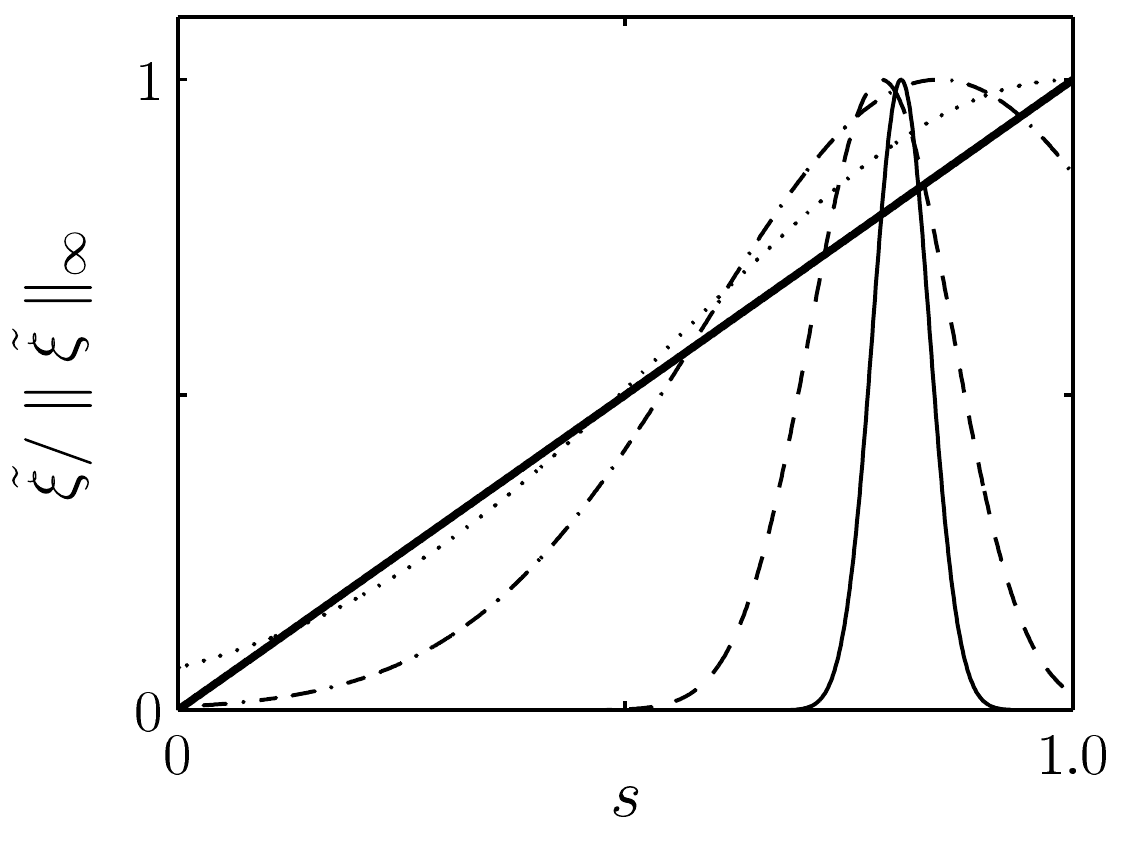}
\label{fig:7b}}
\end{tabular}
}
\end{center}
\caption{Gaussian function optimal configuration: (a) Spread parameter, $\alpha$, (open circles) and centre location, $s_o$, (filled boxes) corresponding to the optimal Gaussian distribution for a given $\xi_0$. (b) Rescaled damping distribution at discrete $\xi_0= 0.004,\ 0.025, 0.25, 0.47$ (solid, dashed, dashed-dot and dotted curves, respectively, indicated on (a) with vertical lines of corresponding description); superimposed is the 
linear optimal distribution ($\xi_1=2$; thick solid curve). {($M^*=12.7,\ U^*=13$)}.}
\label{fig:GaussOpt_distrib}
\end{figure}

Figure~\ref{fig:GaussCon} presents the rescaled power $\mathcal{P}/\mathcal{P}_h^\textrm{max}$ in the $(\alpha,s_o)$-space for small $(\xi_0=0.01)$ and moderate damping $(\xi_0=0.47)$. For small damping, one sees that power is optimally harvested for dampers focused at $s_o=0.8$, the position of the maximum of $\mathcal{K}_0$ along the beam (Figure~\ref{fig:PK_G0}b). This is quite distinct from the high damping optimal, which corresponds to a dispersed distribution ($\alpha=2.7,\ s_o=1$). Of particular note is that for all $\xi_0$, a uniform distribution ($\alpha=0$) is preferred over a concentration of harvesters at $s_o<0.5$. 

We next examine a wider range of damping, $10^{-3}<\xi_0<50$, and plot the optimal values of $(\alpha,s_o)$, in Figure~\ref{fig:7a}. For moderate to large damping, $\xi_0>0.3$, a dispersed distribution with peak at $s_o=1$ is optimal. Conversely for $\xi_0<0.02$, harvesters concentrated at $s_o\approx 0.8$  (position of peak $\mathcal{K}$) is the optimal configuration. Note that for these computations we set $0<\alpha<500$, and at these values of $\xi_0$ the upper bound is reached; more detailed insights and computations on the small damping regime are presented in Section~\ref{sec:sec5}. Figure~\ref{fig:7b} shows the evolution with $\xi_0$ of the optimal normalised distribution and illustrates the transition from focused to dispersed profiles when the total damping is increased.  

\begin{figure}
\begin{center}
\mbox{
\begin{tabular}{c}
{\includegraphics[width=.45\textwidth,angle=0]{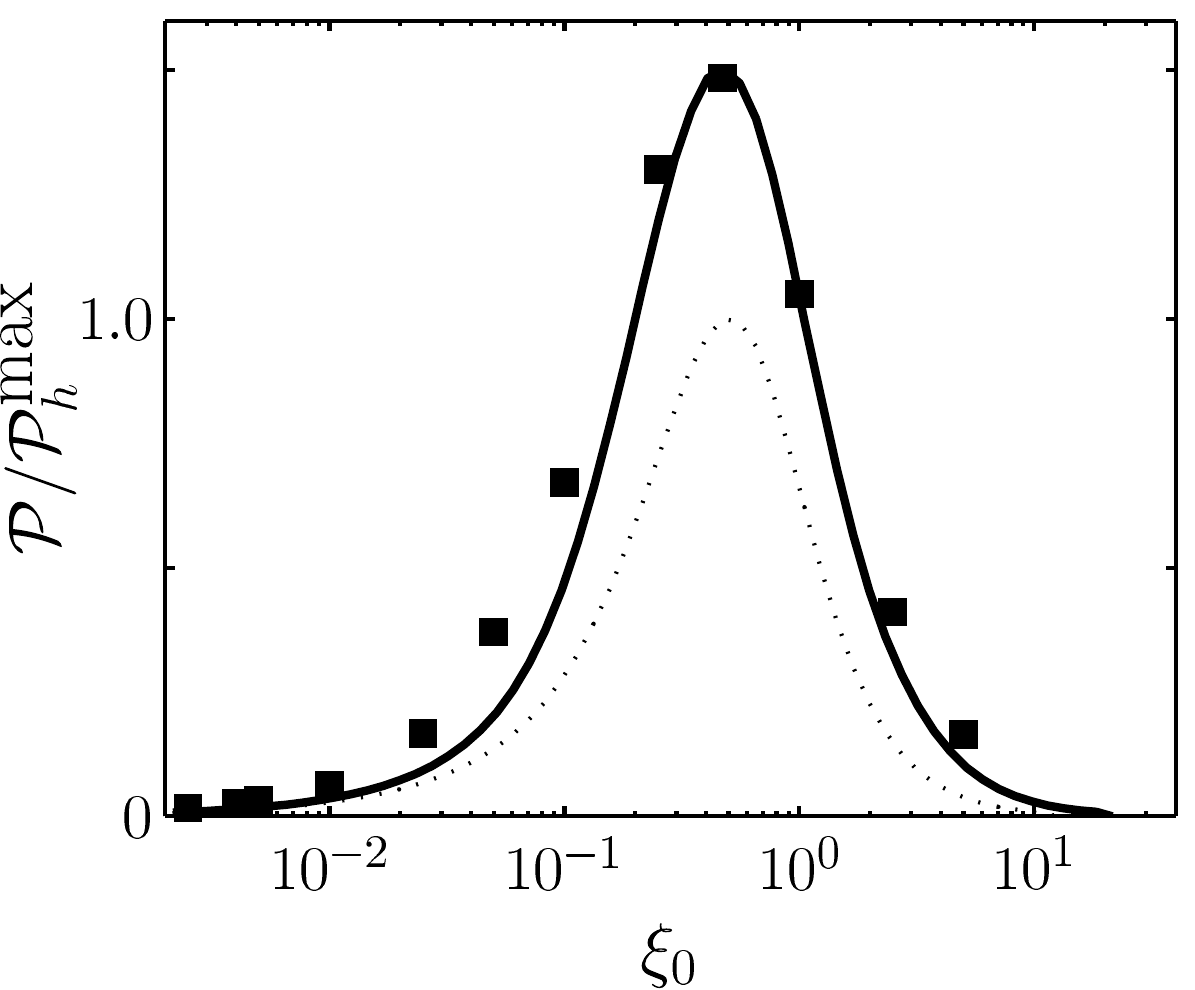}
}
\end{tabular}
}
\end{center}
\caption{Optimal power values corresponding to Gaussian optimal configuration in Figure~\ref{fig:GaussOpt_distrib} (squares) compared to the optimal  power obtained for linear distribution ($\xi_1=2$; thick curve) and constant distribution ($\xi_1=0$; dotted curve).}
\label{fig:GaussOpt_Pow}
\end{figure}

It is possible to determine the optimal harvested power for each value of $\xi_0$; 
in Figure~\ref{fig:GaussOpt_Pow} these optimals are compared with the results for uniform and linear distributions. Strikingly, the peak power values are coincident for linear and Gaussian distributions, which is consistent with the similarity in distributions (Figure~\ref{fig:7b}). Because of the fundamentally different structures of the distribution used, this result suggests that the optimal obtained with such simple functions represents a good approximation of the absolute optimal. Departing from this maximum of harvested power, in particular at small damping we see that concentrated harvesting is superior to the optimal linear distribution.

These results confirm that nonuniform damping distributions can be advantageously employed to enhance the power harvesting capacity. Focused damping distributions are optimal for small damping, whereas dispersed distributions are preferential at higher damping. The following section investigates in more details this transition by considering the impact of damping on the nonlinear dynamical response of the system.

\section{Discussion: Impact of localised damping}
\label{sec:sec5}

The above computations suggest a very different impact of damping on the dynamics of the structure depending on the magnitude of $\xi_0$, thereby leading to quite different optimal strategies to maximise the harvested power. In this section, we first consider the limit of asymptotically small damping before studying the effect of focused dampers on the body's dynamics.

\subsection{Optimal distribution for asymptotically small damping}
\label{sec:loxi}

For asymptotically small damping ($\parallel\xi\parallel_\infty\ll 1$, with $\parallel\xi\parallel_\infty$ the maximum value of $\xi(s)$ for $s\in[0\,\,1]$), the flapping dynamics is not modified at leading order so that $\mathcal{K}=\mathcal{K}_0+O(\parallel\xi\parallel_\infty)$. Thus
\begin{equation}\label{eq:asympt}
\mathcal{P}\sim\frac{1}{M^*U^{*2}}\int_0^1\xi(s)\mathcal{K}_0\ \mathrm{d}s\le \frac{\xi_0}{M^*U^{*2}}\parallel\mathcal{K}_0\parallel_\infty 
\end{equation}
and this upper bound can be approached asymptotically using, for example, the Gaussian distribution \eqref{eqn:Gauss} centred on the maximum of $\mathcal{K}_0$ and increasing $\alpha$. More precisely, any peaked distribution of damping of $L_1$-norm $\xi_0$ centred on this maximum and with a typical width (here $\alpha^{-1/2}$) much smaller than the length scale $\lambda$ associated with the $\mathcal{K}_0$ peak width, should approach the theoretical upper bound $\mathcal{P}^\textrm{th}=\xi_0\parallel\mathcal{K}_0\parallel_\infty/(M^* U^{*2})$, provided that the maximum damping remains small enough that the approximation for $\mathcal{P}$ in Eq.~\eqref{eq:asympt} still holds (which in the present case is equivalent to keeping {$\alpha^{1/2}\xi_0$} bounded).

\begin{figure}
\begin{center}
\mbox{
\begin{tabular}{c}
{\includegraphics[width=.45\textwidth,angle=0]{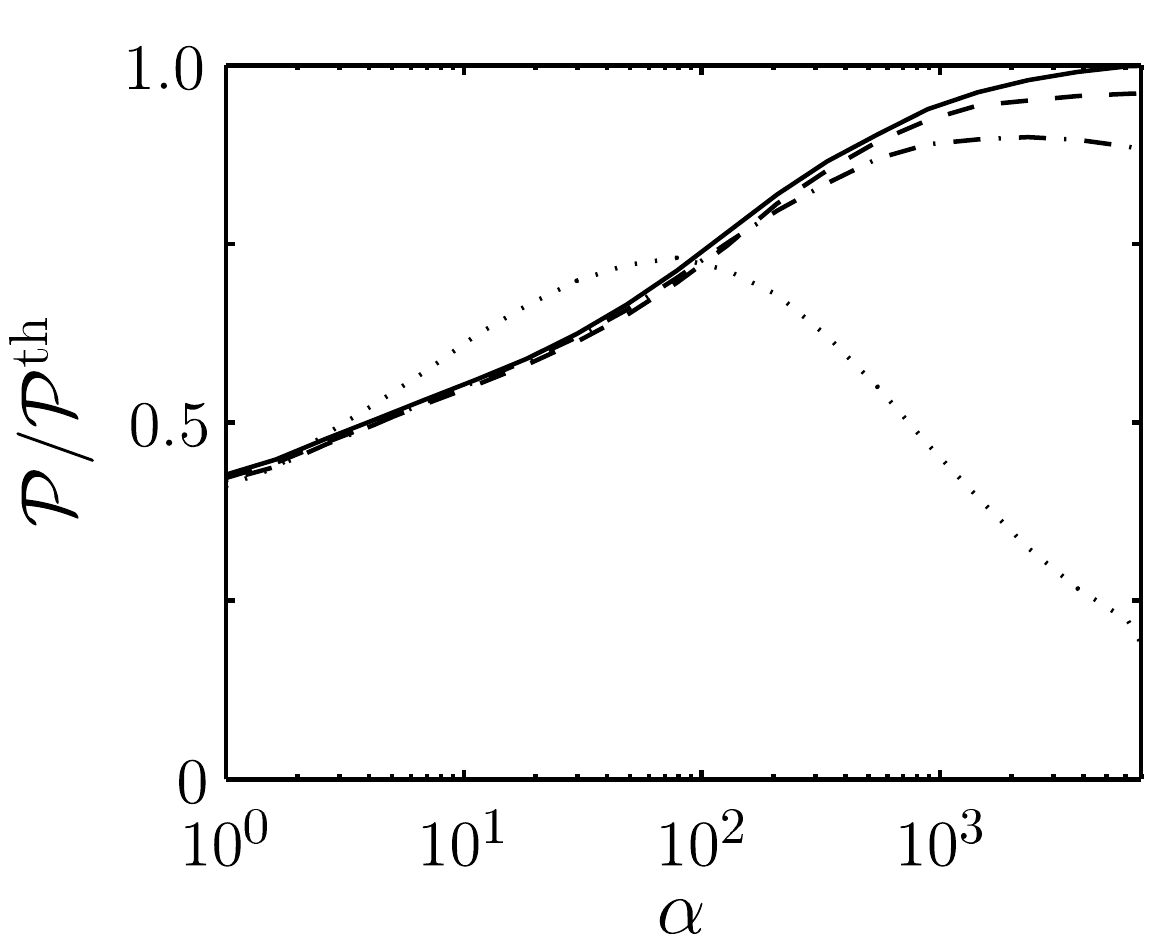}
}
\end{tabular}
}
\end{center}
\caption{Focused harvesters at $s_o=0.81$ for small damping: Curves show the normalised power
for increasingly focused harvesters for small damping values:
  $\xi_0=2.2\times 10^{-5}:10^{-2}$ (solid, dashed, dash-dot, dotted curves, respectively). {($M^*=12.7,\ U^*=13$)}.}
\label{fig:opt_lo}
\end{figure}

This conjecture is tested numerically using a Gaussian distribution, Eq.~\eqref{eqn:Gauss}, with $s_o=0.8$, and by computing the harvested power for small damping: $2.2\times10^{-5}<\xi_0<2.2\times 10^{-2}$. Figure~\ref{fig:opt_lo} shows $\mathcal{P}/\mathcal{P}^{\mathrm{th}}$ and confirms that this conjecture indeed holds since for increasing $\alpha$, the power asymptotically approaches its optimal for small enough $\xi_0$. When $\xi_0\geq 2.2\times 10^{-3}$, the asymptotic value $\mathcal{P}^\textrm{th}$ can not be reached any more: as $\alpha$ is increased, the effect of the maximum damping $\alpha\xi_0$ on the flutter dynamics is important even before the dampers are focused enough for $\mathcal{P}$ to approach $\mathcal{P}^{\mathrm{th}}$.

Two essential results are illustrated here. For sufficiently small damping, the nonlinear response of the structure remains unchanged so the best strategy is to focus all the harvesters in the region of maximum curvature-change. However finite levels of damping significantly modify the nonlinear dynamics of the beam, so at these levels a narrow focusing of damping is sub-optimal; next we investigate further the behaviour at finite damping.

\subsection{Impact of finite and localised damping}
\label{sec:hixi}

\begin{figure}
\begin{center}
\mbox{
\begin{tabular}{cc}
\subfigure[]
{
\includegraphics[width=.42\textwidth,angle=0]{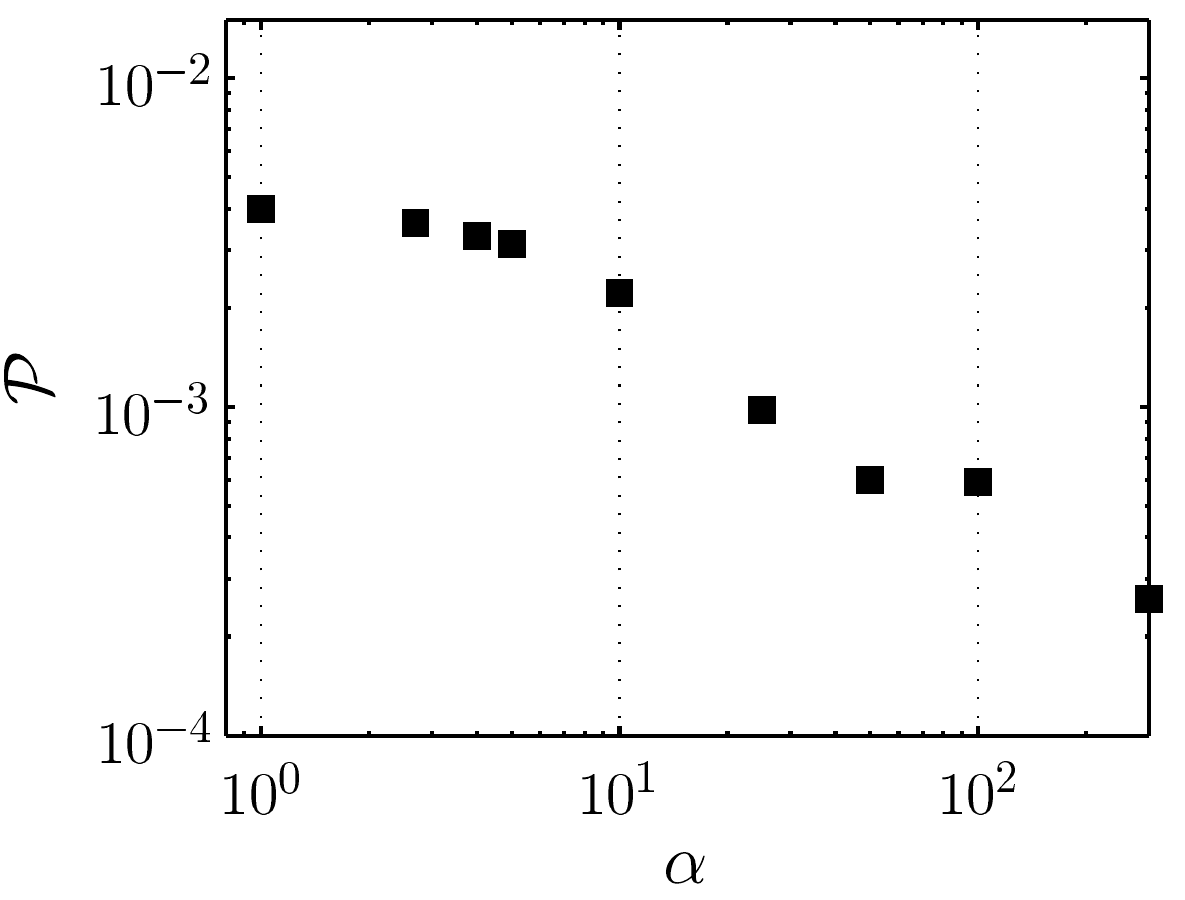}
\label{fig:11a}}
&
\subfigure[]
{
\includegraphics[width=.4\textwidth,angle=0]{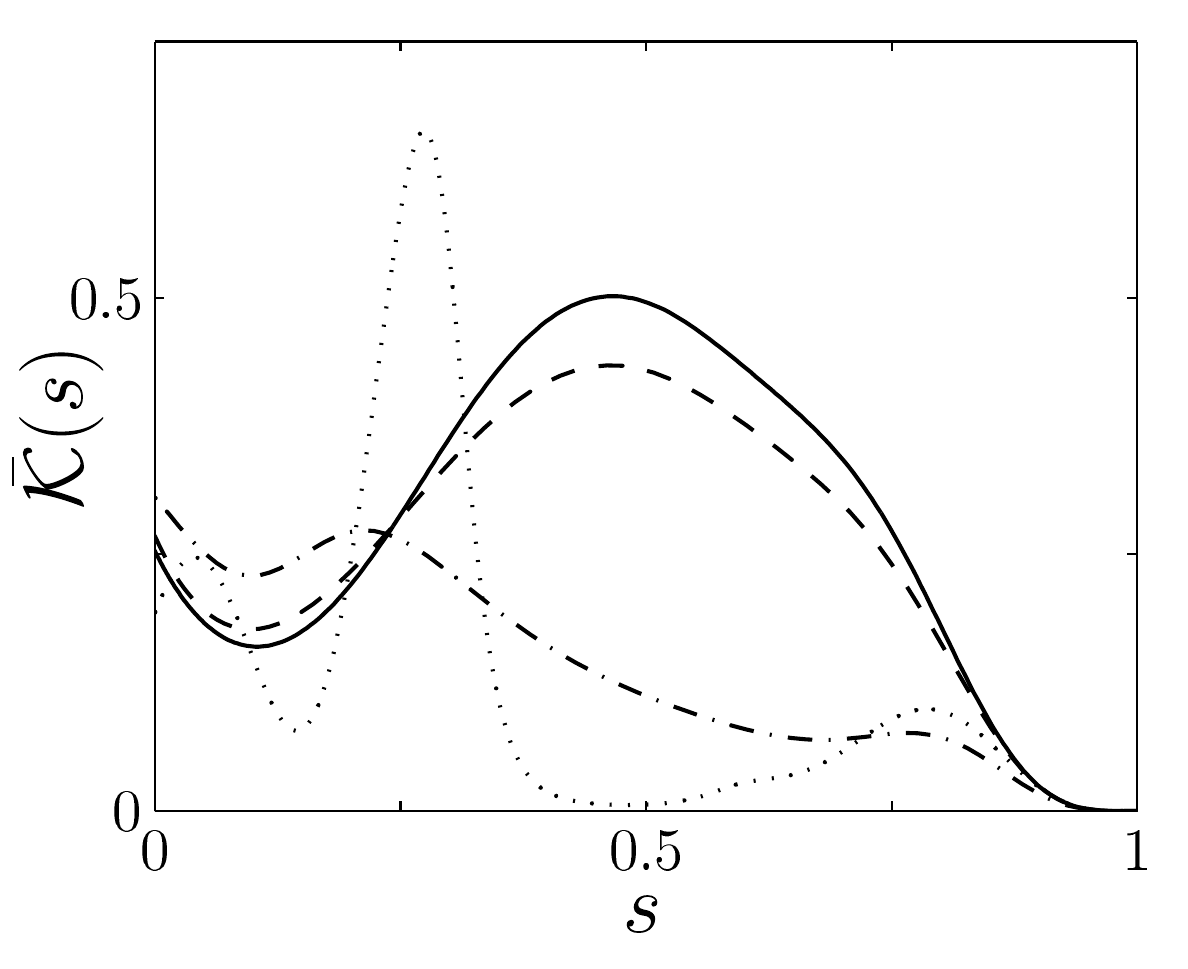}
\label{fig:11b}}
\end{tabular}
}
\end{center}
\caption{Focused harvesters at $s_o=0.45$ for large damping ($\xi_0=0.47$):
 (a) power dependence on $\alpha$
and  (b) $\bar{\mathcal{K}}(s)$ for $\alpha=0,1,10,10^2$
(solid, dashed, dashed-dot, dotted curves, respectively). {($M^*=12.7,\ U^*=13$)}.}
\label{fig:Fig11}
\end{figure}

The computations from Section~\ref{sec:sec4} show that for Gaussian distributions and $\xi_0>0.02$ a defocusing of harvesters is preferable, as the system response becomes increasingly influenced by damping. In order to understand the impact on the system's response, the Gaussian distribution \eqref{eqn:Gauss} is used in the high damping range and $\alpha$ is varied over a range that transitions the distribution from dispersed to focused ($0<\alpha<10^2$) with $\xi_0=0.47$ and $s_o=0.45$ (corresponding to the position of $\parallel \mathcal{K}\parallel_{\infty}$ for $\alpha=0$). Consistent with Figure~\ref{fig:GaussCon}, the harvested power 
in Figure~\ref{fig:11a} decreases with $\alpha$, and Figure~\ref{fig:11b} shows the evolution of $\mathcal{K}(s)$. 

For dispersed distributions ($\alpha < 10$),  $\mathcal{K}(s_o)$ decreases with $\alpha$ and the zone of maximum curvature-change shifts to regions of reduced damping. Nonetheless nonzero damping in this zone is adequate to retrieve some of the energy. For $\alpha > 10$ damping becomes increasingly focused and the flexible body does not deform anymore near the damper ($\mathcal{K}(s_0)\approx 0$): no energy is harvested anymore since no deformation occurs near the harvester's position, and this is reflected in the sharp drop in power in Figure~\ref{fig:11b}. This illustrates the main effect of a focused distribution for finite damping: we see a redistribution of the solid's deformation to regions with little or no damping. This is also the reason why a focused damping distribution is not appropriate for optimal energy harvesting in the finite damping range.

\section{Conclusions}
\label{sec:conc}
In this work, we considered the possibility of harvesting energy from a slender body fluttering in an axial flow, and in particular the impact of harvester-distribution on the performance of the system as well as potential optimisation strategies. To this end, a simplified fluid-solid model was proposed with energy harvesting represented as nonuniform structural damping. Depending on the fluid-to-solid inertia ratio, damping can  actually enhance the flutter response of the structure, as well as reduce the critical flow velocity above which the system can operate. 

For uniform damping distributions, maximising the harvested power appears as a trade-off on the damping intensity: for small damping, the flutter response is only weakly modified and the harvested power increases as more damping is added to the system. For higher damping however, the dynamical response can be strongly modified and the self-sustained oscillations are eventually mitigated. The effect of a nonuniform distribution of harvesters along the structure was considered next. We show that even simple nonuniform distributions such as linear and Gaussian functions, can lead to an increase in harvested peak power on the order of $50\%$. The similarity in the optimal distribution and performance obtained through an optimisation on two fundamentally different families of distributions suggests that simple strategies can capture rather well the characteristics of the global optimal distribution.

Investigating further into the relationship between damping and flutter response, we showed that for small damping, localised harvesting is optimal as it takes full advantage of the system's maximum curvature without impacting its dynamics significantly. On the other hand, for finite damping, focused distributions perform rather poorly as the beam response adapts to rigidify the damped region, leading to negligible harvested power. Instead, nonuniform distributed damping over the entire length of the system becomes optimal.

Returning to the result from the simple bi-articulated model \citep{EH_10}, we note a clear difference in the optimal configurations. Whilst the reduced order model optimal has dampers focused at the fixed end, the continuous optimal requires a dispersed distribution with minimal damping at the fixed end and increasing to a maximum at the free end. However, we find a crucial difference between the two system configurations: whilst the bi-articulated system has a single moving region of curvature (the second articulation) that is responsible for driving the instability, deformations may occur all along the length of the beam in the continuous system. Therefore in the latter, curvature can be displaced away from regions with focused damping while still maintaining the flutter dynamics. In both cases, a careful understanding of the dynamics of the system is necessary to determine the optimal nonuniform positioning of energy harvesters.

\section{Acknowledgements}
The authors gratefully acknowledge the support of Electricit\'e de France (EDF) for their support through the ``Chaire Energies Durables'' at Ecole Polytechnique. S. M. was also supported by a Marie Curie International Reintegration Grant within the $7^\textrm{th}$ European Community Framework Program.

\bibliographystyle{unsrt}
\bibliography{../references.bib}

\end{document}